\RequirePackage[l2tabu, orthodox]{nag}
\RequirePackage{snapshot}

\documentclass[9pt,onecolumn]{extarticle}

\makeatletter
\if@twocolumn
  \usepackage[dvips,letterpaper,top=0.5in, bottom=0.5in, left=0.75in, right=0.5in,includefoot,heightrounded]{geometry}
\else
  \usepackage[dvips,letterpaper,margin=1in,includefoot,heightrounded]{geometry}
\fi

\usepackage{srcltx}

\usepackage[russian,portuges,english]{babel}

\iflanguage{portuges}
    {\newcommand{\keywordname}{Palavras-chaves}}
    {\newcommand{\keywordname}{Keywords}}

\usepackage{amsmath}
\usepackage{amssymb,amsfonts}

\usepackage{abstract}

\usepackage{graphicx}
\usepackage[usenames,dvipsnames,svgnames,x11names]{xcolor}
\usepackage{subfigure}

\usepackage{booktabs}

\usepackage{setspace}
\usepackage{flushend}

\usepackage{cite}

\usepackage{hyperref}
\usepackage[normalem]{ulem}

\usepackage{enumerate}

\usepackage[noend]{algpseudocode}

\usepackage{listings}

\usepackage[short,12hr]{datetime}
       \usepackage{fouriernc}
\makeatletter

\newcommand{\printtitle}{%
\makeatletter
\if@twocolumn

\twocolumn[%
  \maketitle
  \begin{onecolabstract}
    \myabstract
  \end{onecolabstract}
  \begin{center}
    \small
    \textbf{\keywordname}
    \\\medskip
    \mykeywords
  \end{center}
  \bigskip
]
\saythanks
\else
  \maketitle
  \begin{onecolabstract}
    \myabstract
  \end{onecolabstract}
  \begin{center}
    \small
    \textbf{\keywordname}
    \\\medskip
    \mykeywords
  \end{center}
  \bigskip
  \onehalfspacing
\fi
\makeatother
}

\author{%
Bruna G. Palm%
\thanks{%
Universidade Federal de Pernambuco, Recife, Brazil;
Blekinge Institute of Technology, Karlskrona, Sweden.
E-mail: \url{bruna.palm@bth.se}
}
\and
Dimas I. Alves%
\thanks{%
Blekinge Institute of Technology, Karlskrona, Sweden;
Universidade Federal do Pampa, Alegrete, Brazil;
Universidade Federal de Santa Catarina, Florian\'opolis, Brazil.
E-mail: \url{dimasalves@unipampa.edu.br}
}
\and
Viet T. Vu%
\thanks{%
Blekinge Institute of Technology, Karlskrona, Sweden.
E-mail: \url{(viet.thuy.vu, mats.pettersson)@bth.se}
}
\and
Mats I. Pettersson${}^\ddagger$
\and
Fabio M. Bayer%
\thanks{%
Universidade Federal de Santa Maria, Santa Maria, Brazil.
E-mail: \url{bayer@ufsm.br}
}
\and
R. J. Cintra%
\thanks{%
Signal Processing Group, DE/CCEN,
Universidade Federal de Pernambuco, Brazil.
E-mail: \url{rjdsc@de.ufpe.br}
}
\and
Renato Machado%
\thanks{%
Instituto Tecnol\'ogico de Aeron\'autica, S\~ao Jos\'e dos Campos, Brazil.
E-mail: \url{renatomachado@ieee.org}
}
\and
Patrik Dammert%
\thanks{%
Saab Electronic Defence Systems, Sweden.
E-mail: \url{(patrik.dammert, hans.hellsten)@saabgroup.com}
}
\and
Hans Hellsten${}^{\ast\!\!\ast}$
}

\title{%
Autoregressive Model for Multi-Pass SAR Change Detection Based on Image Stacks}

\newcommand{\myabstract}{%
Change detection is an important synthetic aperture radar (SAR) application, usually
used to detect changes on the ground scene measurements
in different moments in time.
Traditionally,
change detection algorithm (CDA)
is  mainly
designed for two synthetic aperture radar (SAR) images retrieved at different instants.
However, more images can be used to improve
the algorithms performance, witch emerges as a research topic on SAR change detection.
Image stack information can be treated as a data series over time
and can be modeled by autoregressive (AR) models.
Thus, we present some initial findings on SAR change detection based on image stack considering
AR models.
Applying AR model
for each pixel position in the image stack,
we obtained an estimated image of the ground scene which can be used as a reference image for CDA.
The experimental results reveal that ground scene estimates by the
AR models is accurate and can be used for change detection applications.
}

\newcommand{\mykeywords}{%
AR models, change detection, SAR, time series.
}

\date{}

\begin{document}

\printtitle

\section{Introduction}

Autoregressive (AR) models
can be
used to describe
random processes,
for example
time-varying processes in nature.
The output of an AR model is a prediction
based on the
temporal statistical
characteristics
of
a given time-varying process~\cite{ghirmai2015,Box2008}.
AR models are used in
statistics and signal processing
applications,
such as in~\cite{ghirmai2015,liu2014,biscainho2004,milenkovic1986}.
In this study,
an AR model is considered
in the context of
synthetic aperture radar (SAR)
change detection.

The SAR change detection algorithm (CDA)
refers to
the methods used to
detect changes in a ground scene between distinct
measurement
in time.
The changes on the ground scene can be the result of
either
natural disasters like flood and wildfire
or man-made interference,
such as
deforestation
and
installations~\cite{ulander2004,folkesson2009,ulander2006}.
SAR change detection is a common method
of detecting
changes
in SAR images (reference and surveillance)
over the same area,
but at different instants.
The requirements for such
measurements are that the passes,
the heading
angles of the platform, and the incident angles,
are almost identical~\cite{Renato2016}.

The idea of multi-pass SAR change detection based
on image stack was brought up recently~\cite{vu2017}.
The idea comes from the fact that with two (or more) reference radar
images instead of one, more knowledge
on ground clutter can be retrieved.
This knowledge is used to eliminate
clutter and noise in the surveillance image that
can enhance CDA results.
One of the initial study on this topic
is presented in~\cite{vu2017}, where a small stack
with three images is experimented.
The aim is to minimize the false alarm rate
caused by
the clutter especially
when elongated
structures such as power lines and fences
are considered
in the SAR scene.
Two out of three images without change are used to calculate the statistics of the clutter.
The surveillance image with changes is placed into the
output of an adaptive noise canceler (ANC)
and
the
difference image
is used as a reference signal.
Experimental results indicate that the process with small image stacks
can reduce significantly the incidence of false alarms and provide
a high
probability of detection.
Based on the same stack size, a change detection method is
proposed in~\cite{vu2017}.
The method is based on the analysis
of the distribution
obtained by
the
differences between two images without change
(simply obtained by a subtraction)~\cite{Renato2016}.
As shown in~\cite{Renato2016},
the distribution is very close to a Gaussian process.
A likelihood ratio test (LRT) based on
the
Neyman-Pearson lemma can be easily obtained with the
bivariate Gaussian probability density function
which
plays
a significant role
in
the proposed change detection method~\cite{vu2017}.

For large image stacks, composed of more than three images,
and based on the assumption that the heading
angles of the platform and the incident
angles of the multi-passes are identical,
not only the knowledge on ground clutter
is retrieved. In theory,
the whole ground scene can be predicted accurately.
By using an AR model,
it
is possible to
describe
the changes in the measurements that occur in the ground scene over time.
Thus, it is possible to get the
prediction of the true ground scene
without change. Once the prediction of
the ground scene without change is available,
the changes in the ground scene can be easily detected.
In this paper, we present an initial study
on ground scene prediction by using
AR models for SAR image stacks.
The data used in this study
consist of
eight SAR images obtained by the CARABAS II system~\cite{Lundberg2006}.

The paper is organized as follows.
Section~\ref{s:series}
introduces the  time series model considered in this
study.
Section~\ref{s:tec} details the
ground estimation
technique considered in this paper.
Section~\ref{s:res}
presents the data description and experimental results.
Concluding remarks are in Section~\ref{s:con}.

\section{Autoregressive model}
\label{s:series}

Let~$y[n]$
be an~$N$-point
time-series indexed
over
$n=1,2,\ldots , N$.
With the analysis of a stationary series,
it is possible to extract information on
time series data, e.g., statistics;
whereas time series forecasting
uses
time series models
to predict the future data
based on the previous data~\cite{Brockwell2016}.
In this study, the AR model for time series
is considered,
which can be expressed by~\cite{Kay1993}
\begin{align}
\label{e:model}
y[n] = - \sum_{k=1}^p a[k] y[n-k] + u[n]
,
\end{align}
where~$y[n]$ is the amplitude value
of each pixel in one image,
$a[k]$
are the autoregressive terms,
$u[n]$ is white noise,
and
$p$~is the order of the model~\cite{kay1998}.
The autoregressive terms~$a[k]$ in Equation~\eqref{e:model}
can be estimated by the Yule-Walker estimator~\cite{Brockwell2016}.
Hence, the estimated autoregressive terms~$\widehat{a}[k]$
are the solutions of the following equation system
\begin{align*}
\begin{bmatrix}
r_{yy}[0] & r_{yy}[1] & \ldots & r_{yy}[p-1] \\
r_{yy}[1] & r_{yy}[0] & \ldots & r_{yy}[p-2] \\
\vdots & \vdots & \ddots & \vdots \\
r_{yy}[p-1] & r_{yy}[p-2] & \ldots & r_{yy}[0]
\end{bmatrix}
\begin{bmatrix}
a[1] \\ a[2] \\ \vdots \\ a[p]
\end{bmatrix}
= -
\begin{bmatrix}
r_{yy}[1] \\ r_{yy}[2] \\ \vdots \\ r_{yy}[p]
\end{bmatrix}
,
\end{align*}
where~$r_{yy}[\cdot]$ is the autocorrelation
function.
The information about large sample distributions of the Yule-Walker
estimator, order selection, and confidence
regions for the coefficients can be found in~\cite{Brockwell2013}.

Considering the
estimated autoregressive terms,
it is possible to forecast~$h$ steps
forward with the AR model as:~\cite{Brockwell2016}
\begin{align}\label{e:prev}
\widehat{y}[N+h] = - \sum \limits _{k=1} ^p
\widehat{a}[k] \widehat{y}[N+h-k].
\end{align}

In the next section, we use
the AR model for a stack of eight
SAR images
to forecasting the ground
reference image.

\section{Ground estimation for change detection}
\label{s:tec}

Basically, CDA
requires two images (reference and surveillance) associated with
exactly two measurements.
The processing sequence includes change detection
and change
classification or false alarm minimization.
Conceptually,
change detection
can be simply obtained by a
subtraction of the reference image with the surveillance image,
followed by thresholding.

However, recent studies have shown
that image stacks can be used to improve the performance of change
detection~\cite{vu2017}.
The stack should be
composed by
images with the same
heading and
incident angle.
In this study we consider
a stack composed
of eight images
with the same ground scene
with four different targets positions.

In the signal processing point of view,
an image stack can be viewed as a time series, since
it is a series of data indexed over the time.
It is natural to apply time series
analysis and time series forecasting for
information extraction and obtain prediction data.
With the goal of change detection,
the information extraction can concern the
statistics of SAR image stack while the
information prediction can be the ground scene
estimation.

\begin{figure}
\centering
\includegraphics[scale=0.25]{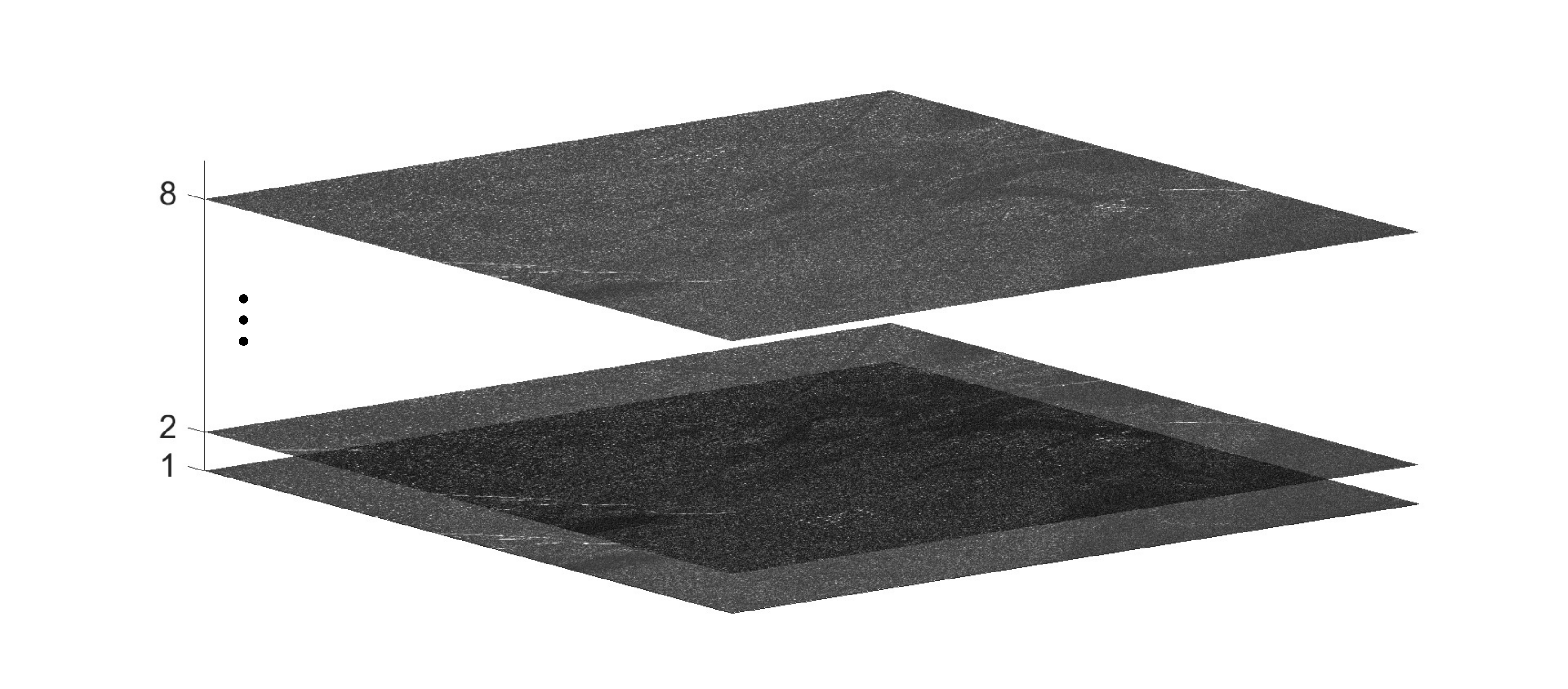}
\caption{Images samples of the stack. The stack is used to predict
a ground scene reference image.}\label{f:pac}
\end{figure}

Based on the AR model presented in the previous section,
it is possible to obtain a ground estimation of the images.
The used images in the stack are assumed to be
perfectly
co-registered.
The image samples extracted from the
stack of images
are illustrated in Figure~\ref{f:pac}
and they
form a time series that is placed into an AR model.
The extraction and prediction are applied to
the pixel positions.
As a result, we obtain a true image of the
ground scene illuminated by multi-pass.

\begin{figure}
\centering
\includegraphics[scale=0.27]{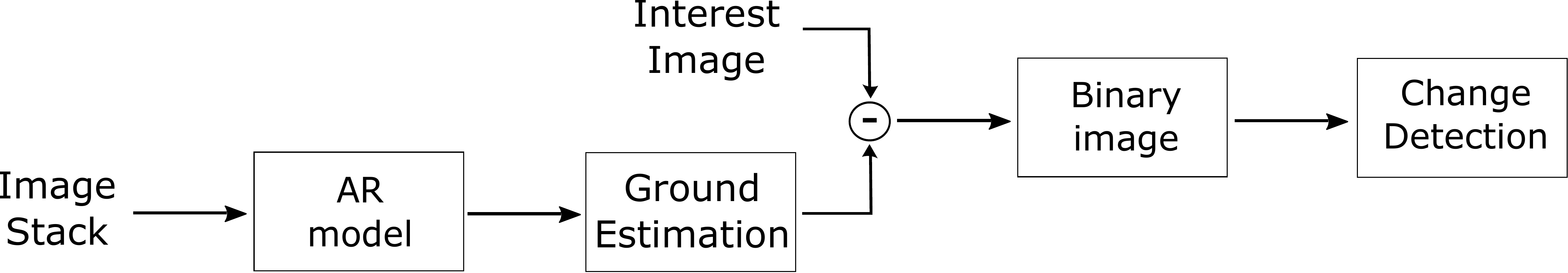}
\caption{Processing scheme for change detection.}
\label{f:proc}
\end{figure}

Once the predicted image is available,
it can be used, for example, as a reference
image for change detection. In this case, the
changes of the ground scene before the first
pass and the latter passes can be detected.
To evaluate this proposal, we performed a
simple processing sequence for the stack
of images including ground estimation,
change detection and change classification, as shown
in Figure~\ref{f:proc}.

The stack of images is used as input
of the AR model, which provides a ground
reference image. Then,
each image
used in the stack is
compared to the
reference image,
resulting in difference images.
The processing is
followed by thresholding given rise to a binary image.
The last processing step is reserved for
morphological operations such as erosion and dilation.
For the experiments, we performed
an erosion followed by a dilatation.

The measures of performance of
change detection
is based on the
probability of detection~($\text{P}_d$) and
false alarm rate (FAR).
The quantity~$\text{P}_d$
is obtained
from the ratio between the
number of
detected targets and the total
numbers of known targets,
and FAR is defined
by the number of false
alarms
detected
per square kilometer~\cite{Lundberg2006}.

\section{Experimental results}
\label{s:res}

In this section we provide some
experimental results
to assess the performance obtained by
the processing scheme presented in Section~\ref{s:tec} by considering
predicted images as ground estimation input.

\subsection{Data description}

\begin{figure}
\centering
\subfigure[Mission  one]
{\label{f:im1}\includegraphics[width=0.2\textwidth]{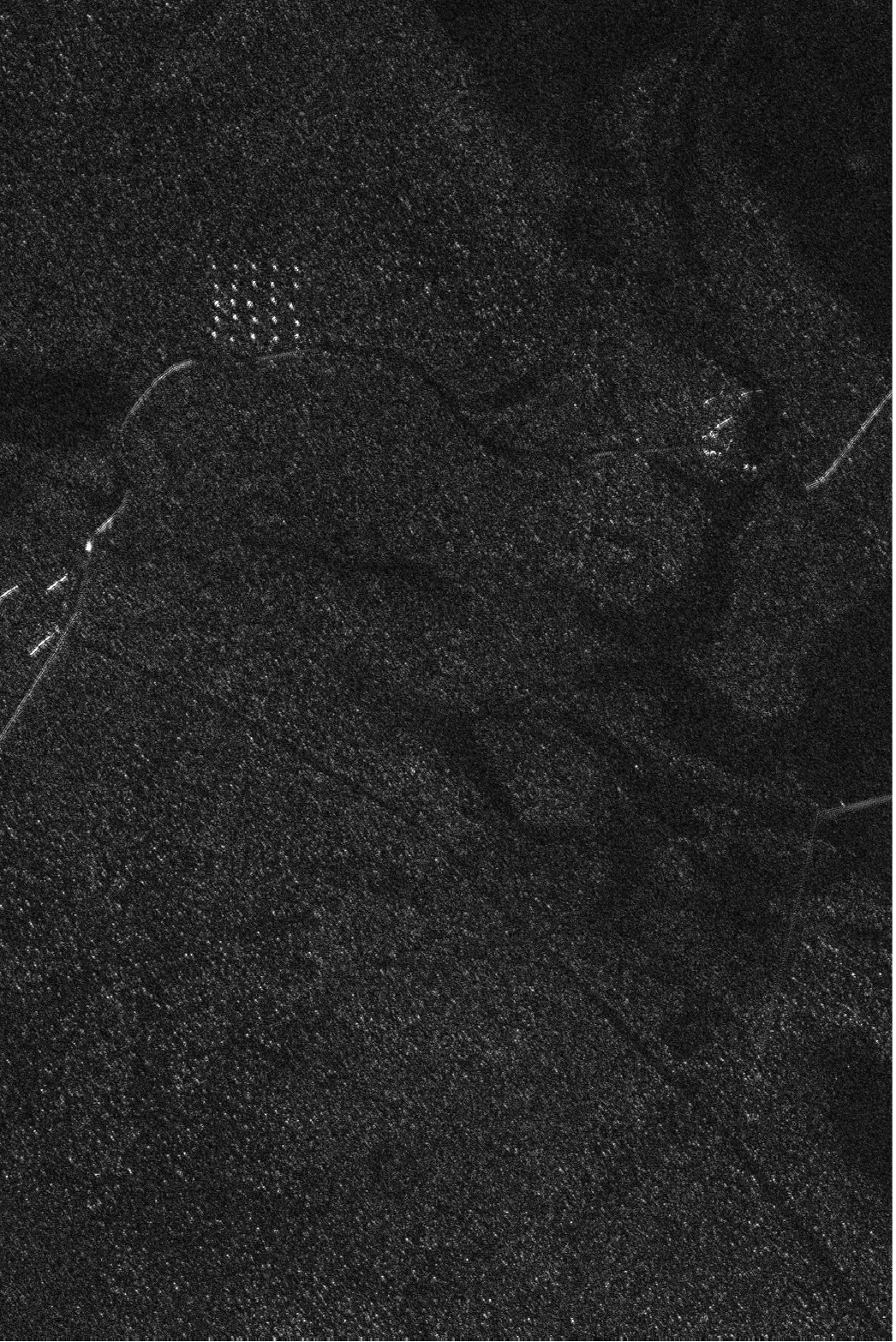}}
\subfigure[Mission  two]
{\label{f:im2}\includegraphics[width=0.2\textwidth]{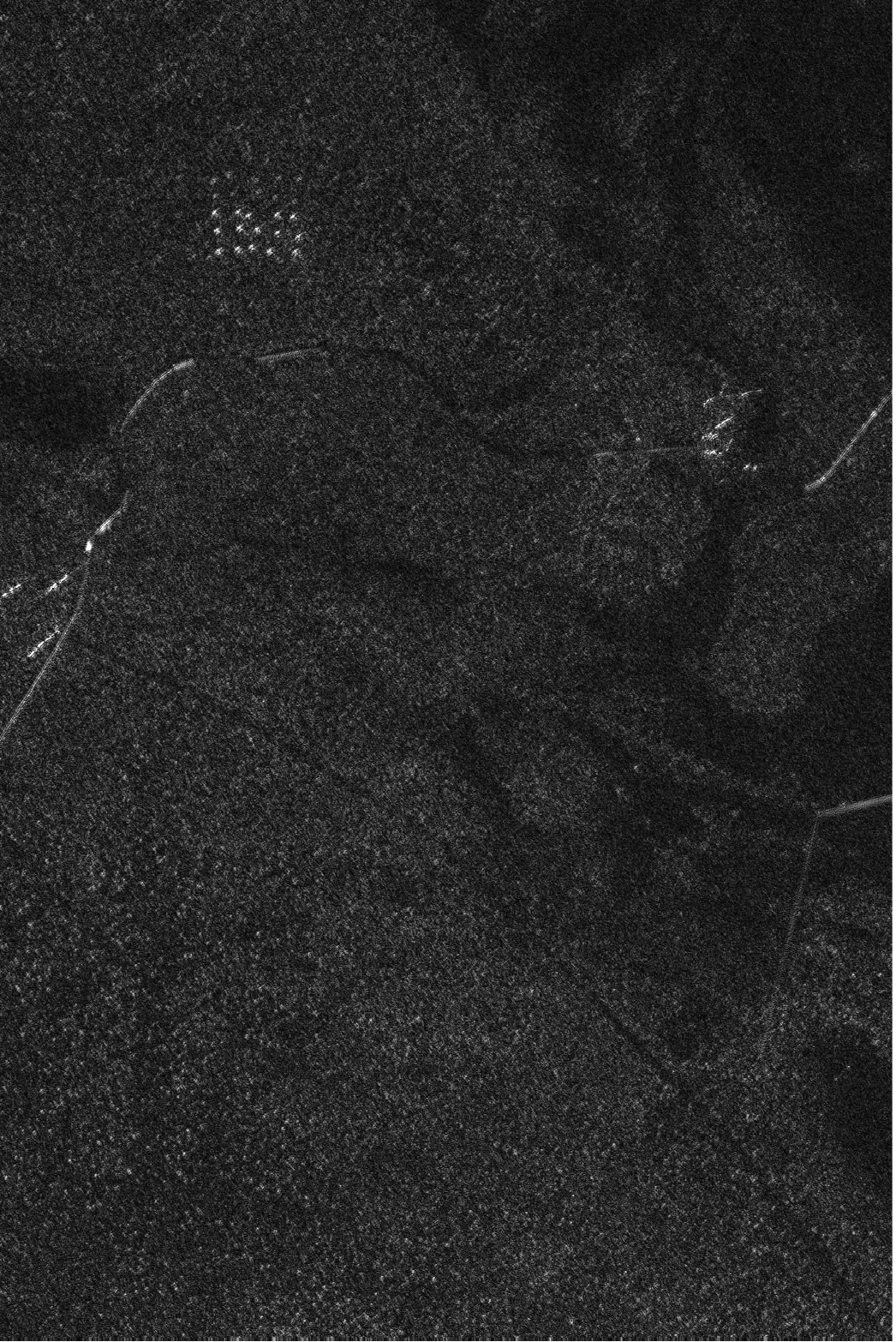}}
\subfigure[Mission three ]
{\label{f:im3}\includegraphics[width=0.2\textwidth]{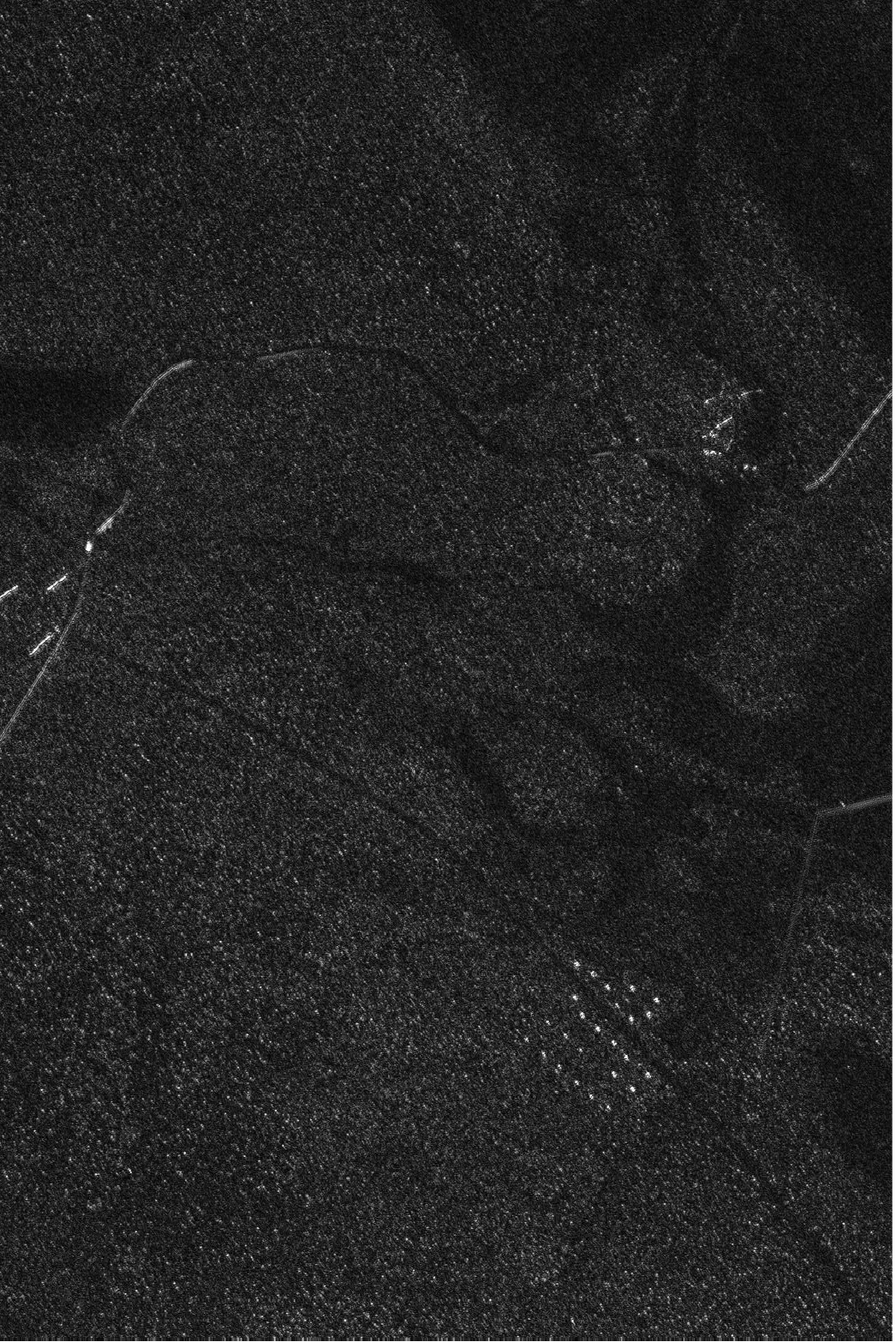}}
\subfigure[Mission  four]
{\label{f:im4}\includegraphics[width=0.2\textwidth]{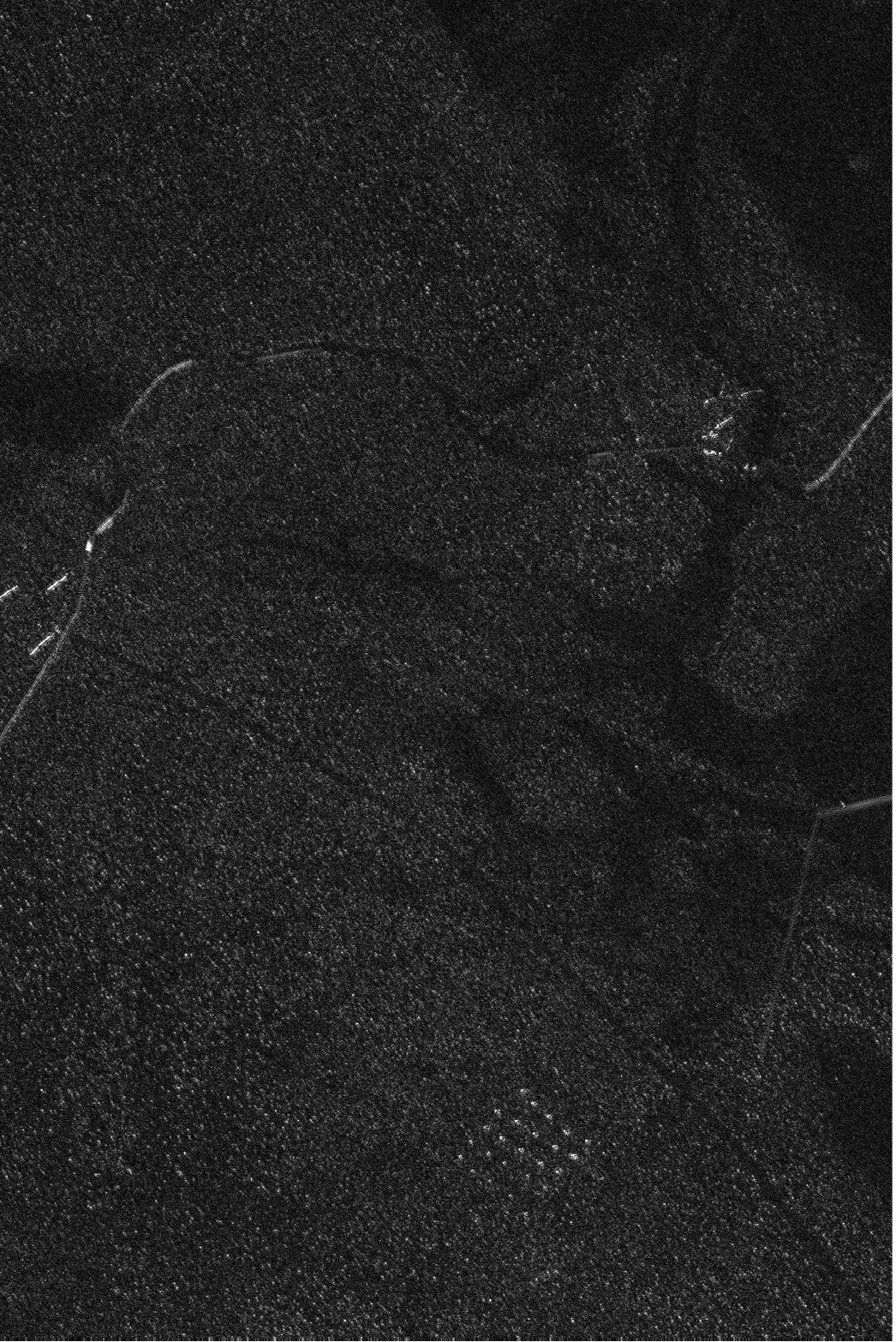}}
\caption{CARABAS II images for pass five.}\label{f:pass5}
\end{figure}

\begin{figure}
\centering
\subfigure[Mission  one]
{\label{f:im5}\includegraphics[width=0.2\textwidth]{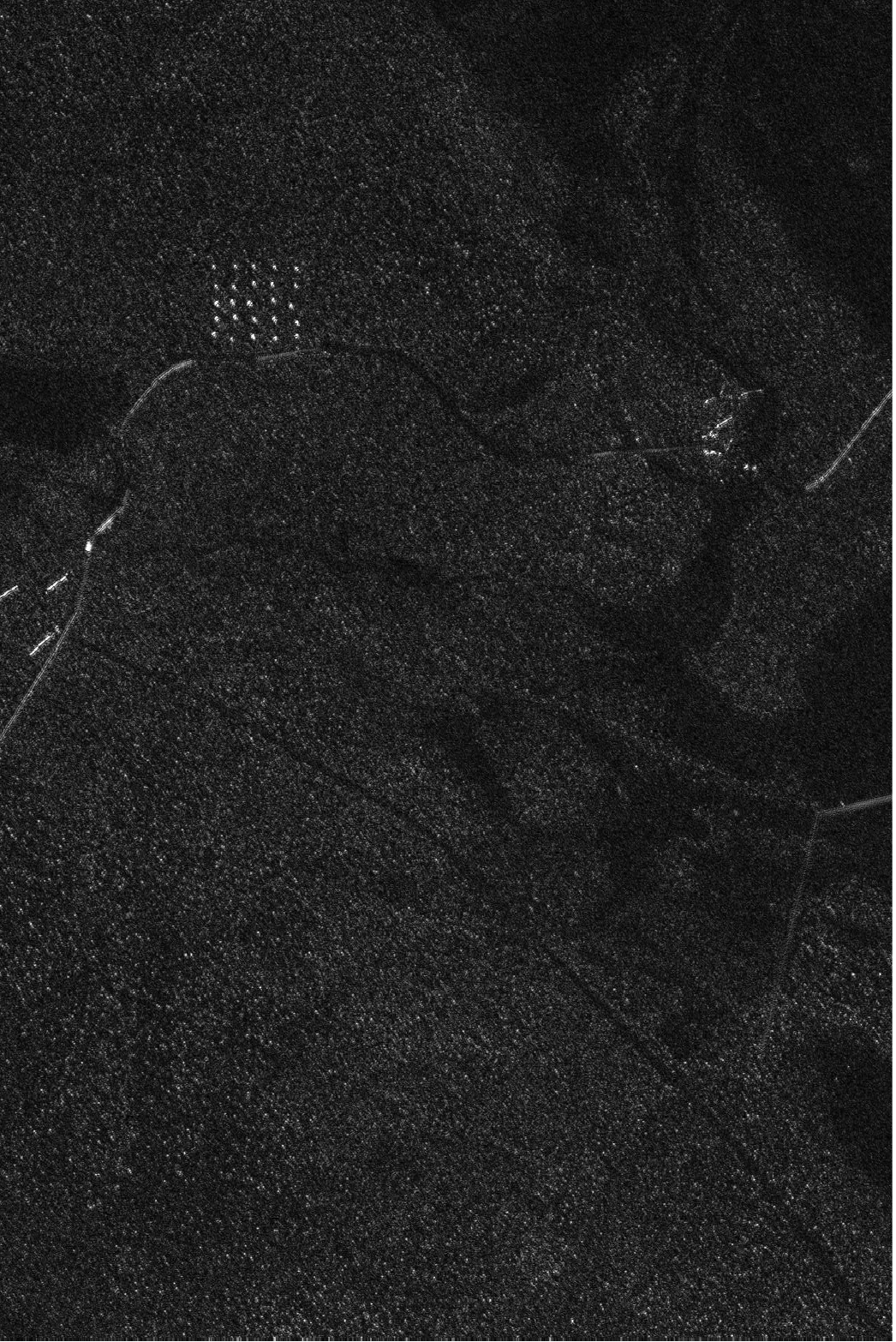}}
\subfigure[Mission  two]
{\label{f:im6}\includegraphics[width=0.2\textwidth]{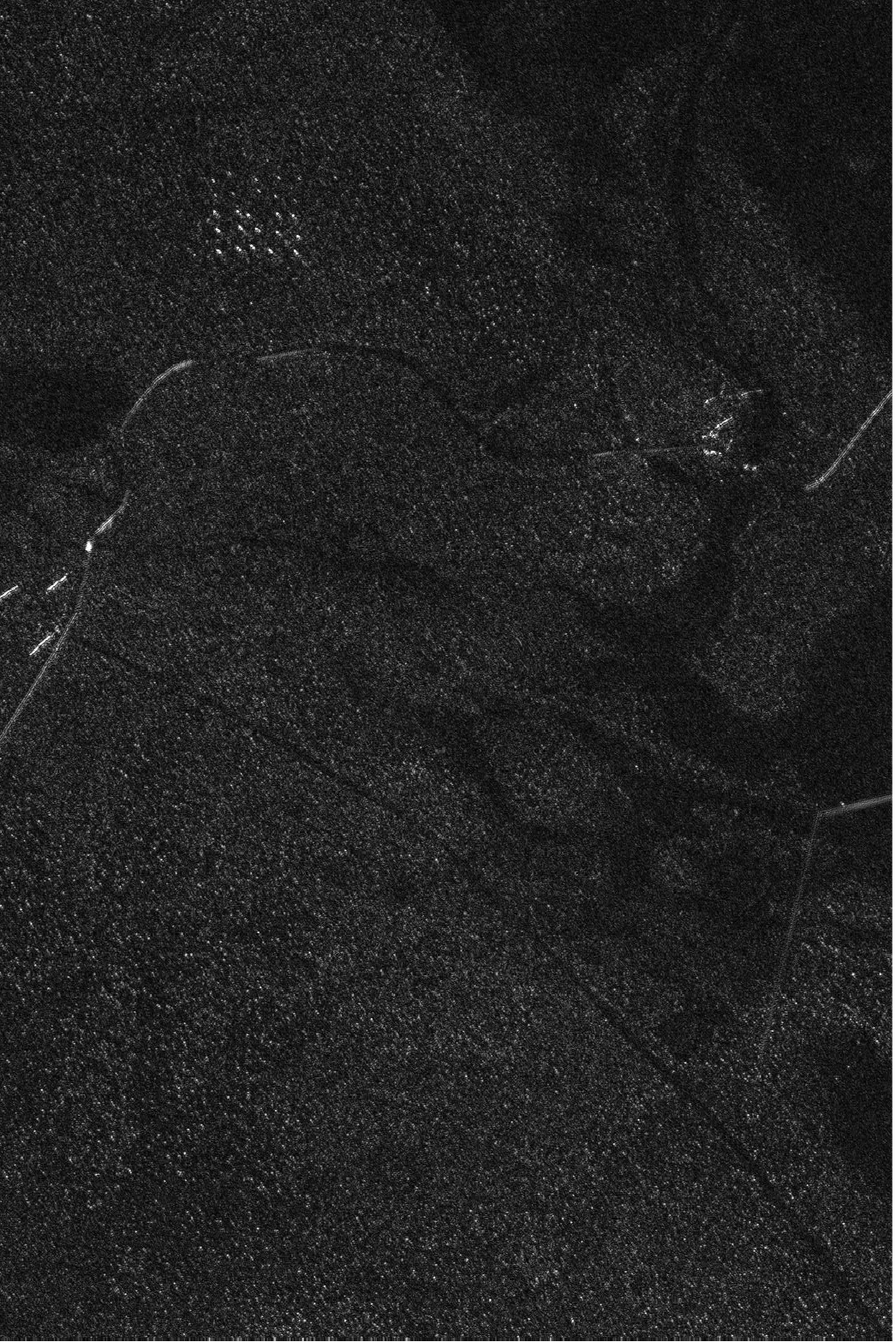}}
\subfigure[Mission three]
{\label{f:im7}\includegraphics[width=0.2\textwidth]{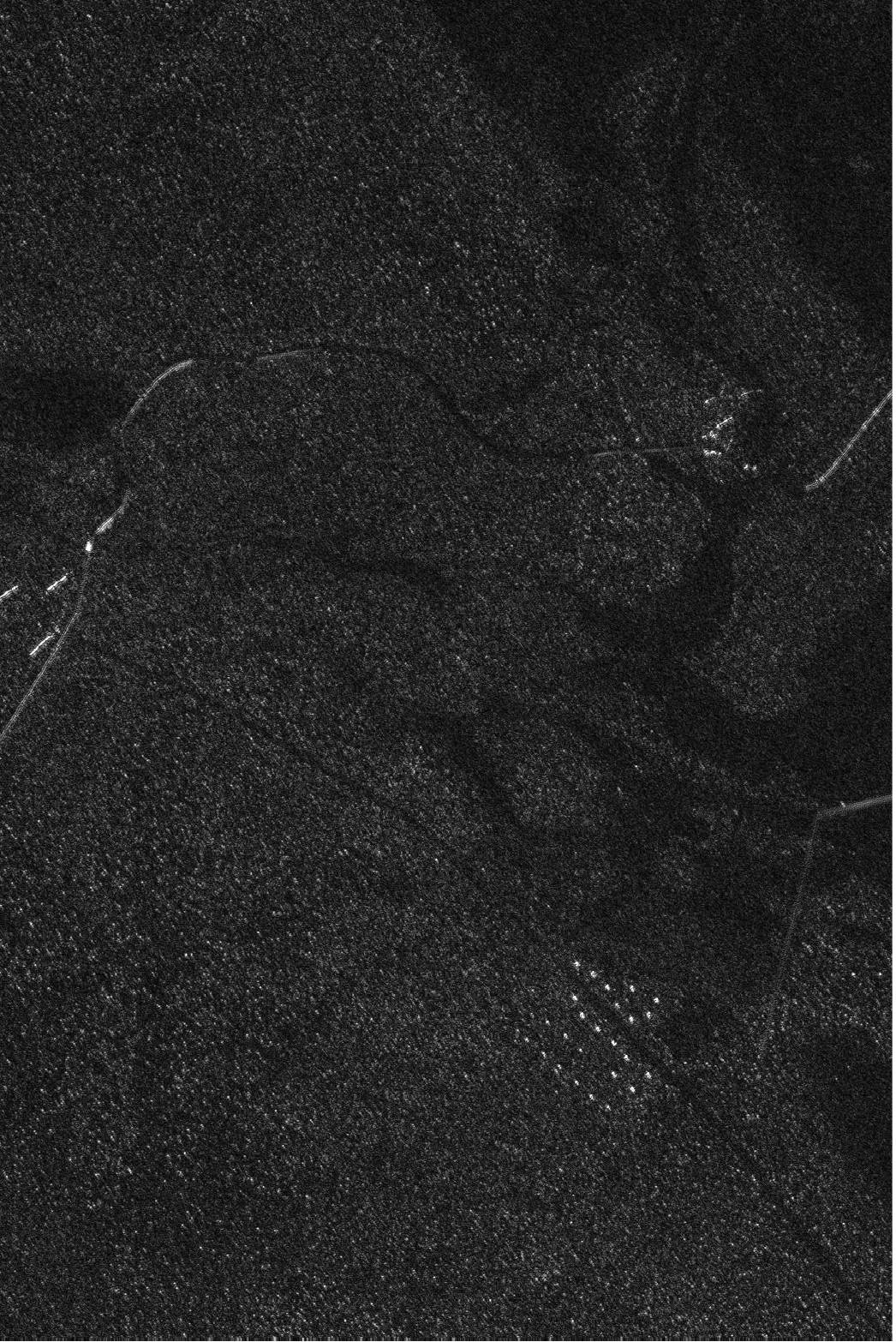}}
\subfigure[Mission  four]
{\label{f:im8}\includegraphics[width=0.2\textwidth]{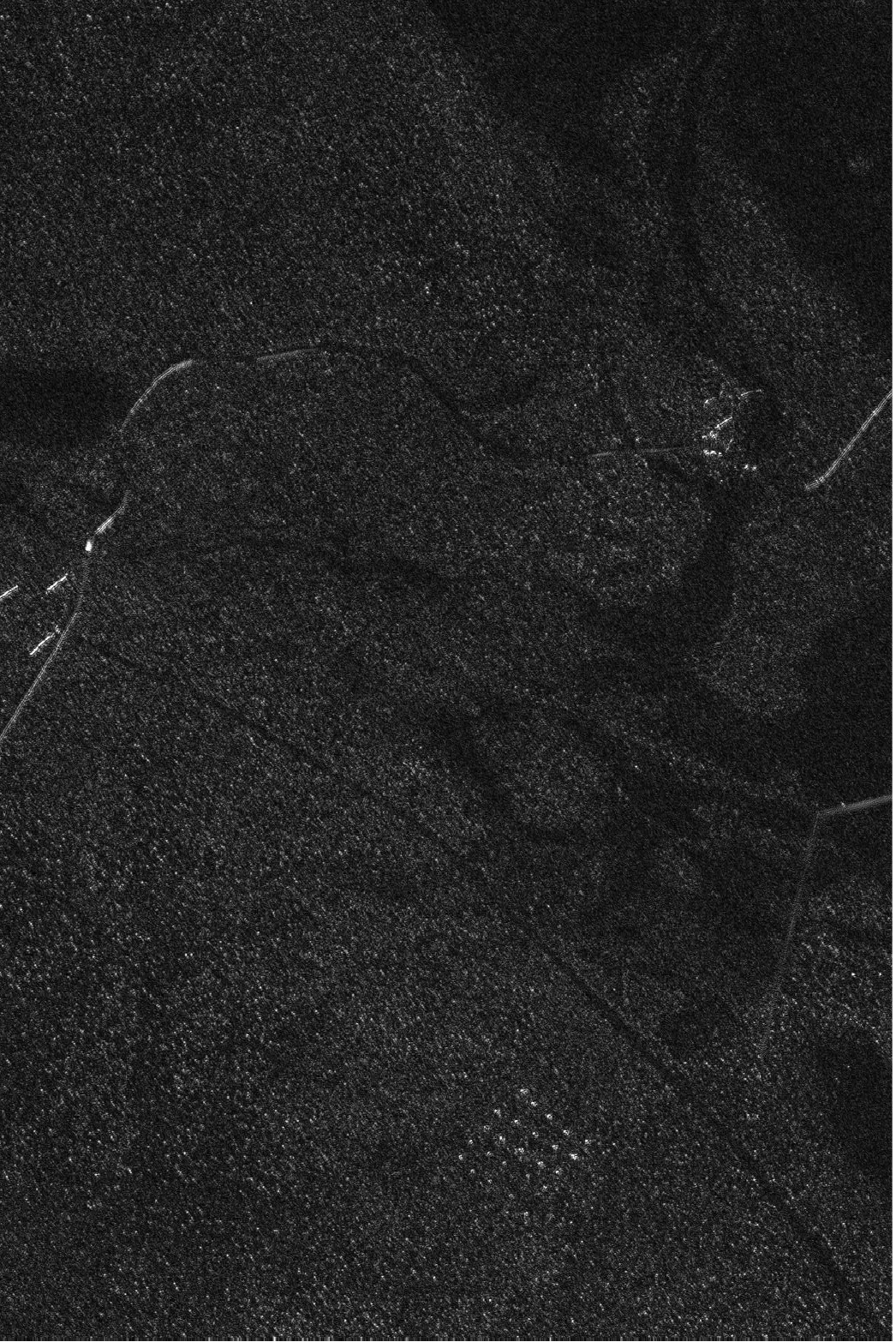}}
\caption{CARABAS II images for pass six.}\label{f:pass6}
\end{figure}

The data used for this study were delivered by
CARABAS~II~\cite{Lundberg2006},
a Swedish ultra-wideband VHF SAR system.
The data included eight
images with almost identical
flight geometry,
but
with
four different targets deployments in the ground
scene~\cite{Ulander2005,Lundberg2006}.
All information about the data
can be found in~\cite{Ulander2005,Lundberg2006}.
In this paper, we consider
the image stack
generated by
passes five and six
and missions one to four
for each pass.
Figures~\ref{f:pass5}
and~\ref{f:pass6}
present eight
CARABAS~II
images forming the image
stack
used in
the AR model.
Each image is
represented by
a matrix of
$3000 \times 2000$
pixels, corresponding
to
an area of~$6~\text{km}^2$.
The ground scene is
dominated by forest with pine trees.
Fences, power lines and roads were also present in the scene.
Some military vehicles were deployed in the SAR scene and
placed in a manner to facilitate
their identifications in the tests~\cite{Lundberg2006}.
The targets
can be seen
in the
upper-left
corner in
Figures~\ref{f:im1},~\ref{f:im2},~\ref{f:im5},
and~\ref{f:im6};
and in the lower-right corner
in
Figures~\ref{f:im3},~\ref{f:im4},~\ref{f:im7},
and~\ref{f:im8}.
In Figures~\ref{f:pass5}
and~\ref{f:pass6}
it is possible
to identify
the targets
and other structures,
like roads and
power lines.
Each image
has $25$~targets
with
three
different
sizes~\cite{Lundberg2006}.


\subsection{Ground estimation}

\begin{figure}
\centering
\includegraphics[scale=0.7]{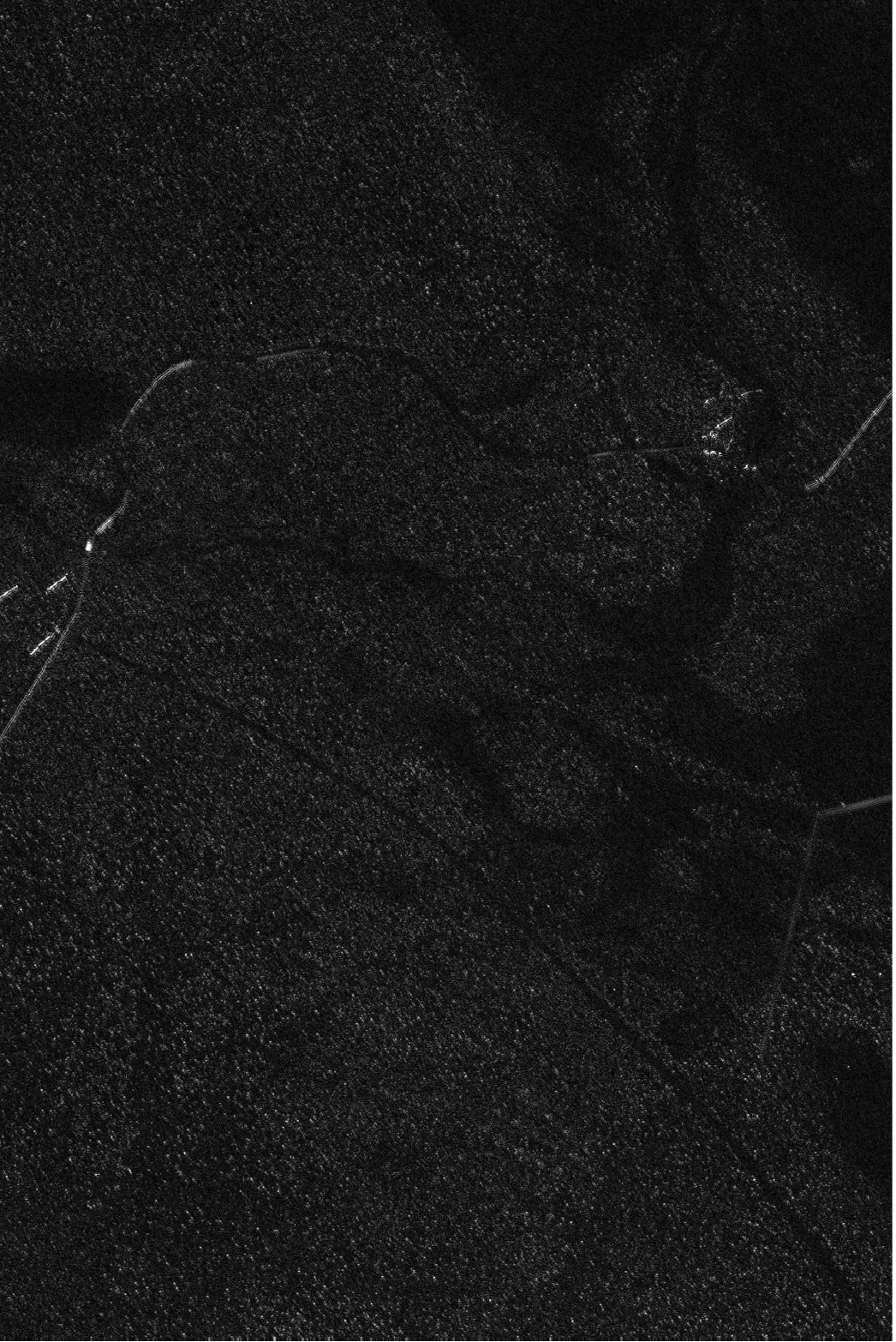}
\caption{Prediction image of the AR$(1)$ model.}\label{f:prev}
\end{figure}

In this study we considered
the AR model
presented in Section~\ref{s:series}.
The resolution of the CARABAS~II system
is approximately
$3 \times 3 \; \text{m}^2$.
Since a pixel
represents
a
$1 \times 1 \; \text{m}^2$,
considering an
one-dimensional model,
the closest pixels will be more correlated
than the others.
Thus, we used~$p=1$
in the AR models.
Based on the fitted model,
we obtained the forecast of
one step ahead for each pixel,
like showed in
Equation~\eqref{e:prev}.
This forecasting provides
a new image presented
in Figure~\ref{f:prev}
representing
the ground estimation
of the SAR scene.

We can visually compare the images considered
for the ground estimation
(showed in
Figures~\ref{f:pass5}
and~\ref{f:pass6}),
and the ground estimation
(the new image
presented
in Figure~\ref{f:prev}).
Hence, we can verify the lake regions, the forest, and strong
scatters in all images.
However, the deployed vehicles
in the missions, that can be verified in different locations
in Figures~\ref{f:pass5}
and~\ref{f:pass6}, do not appear in Figure~\ref{f:prev}.
With visual verification, Figure~\ref{f:prev} seems to present
a good estimation of the ground scene.

\begin{figure}
\centering
\includegraphics[scale=0.7]{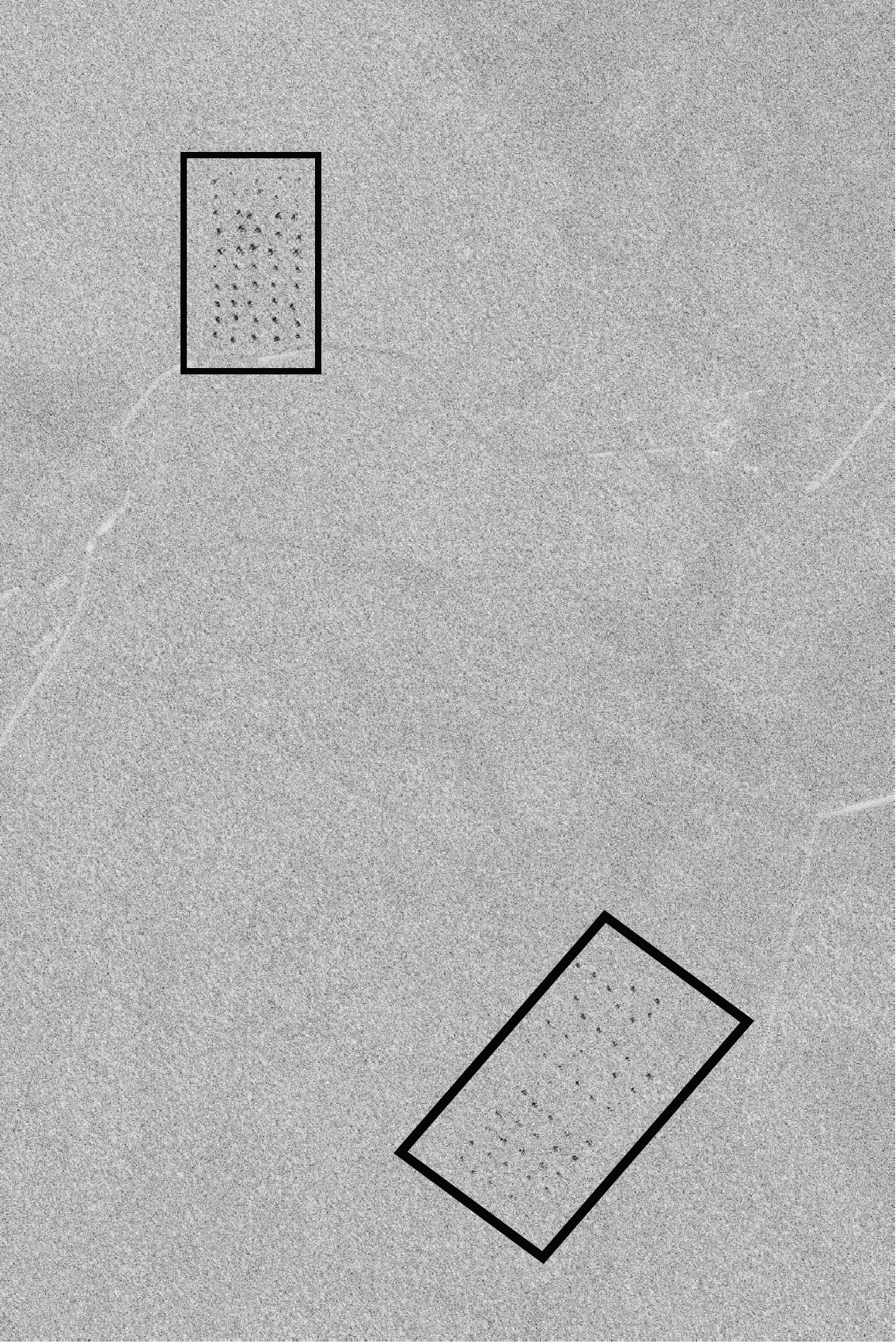}
\caption{Absolute values of the autoregressive terms estimated by the AR$(1)$ model.
The black points correspond to the lowest values.}\label{f:coef}
\end{figure}

Figure~\ref{f:coef} presents the
image with the
absolute estimative values
of the
autoregressive terms
of the AR$(1)$ model.
The black points
in the image correspond
to the lowest values.
It is possible
to see
(in the
boxes)
the
regions
where the vehicles were deployed in
some instants in time
more highlighted
than the other structures in the image.
Thus, the
AR model
reduce
the
amplitude
of the targets
in the ground
estimation.
In future studies,
this image can
be used to indicate the changes
in the ground scene.

\begin{figure}
\centering
\subfigure[Mission one]
{\label{f:res1}\includegraphics[width=0.2\textwidth]{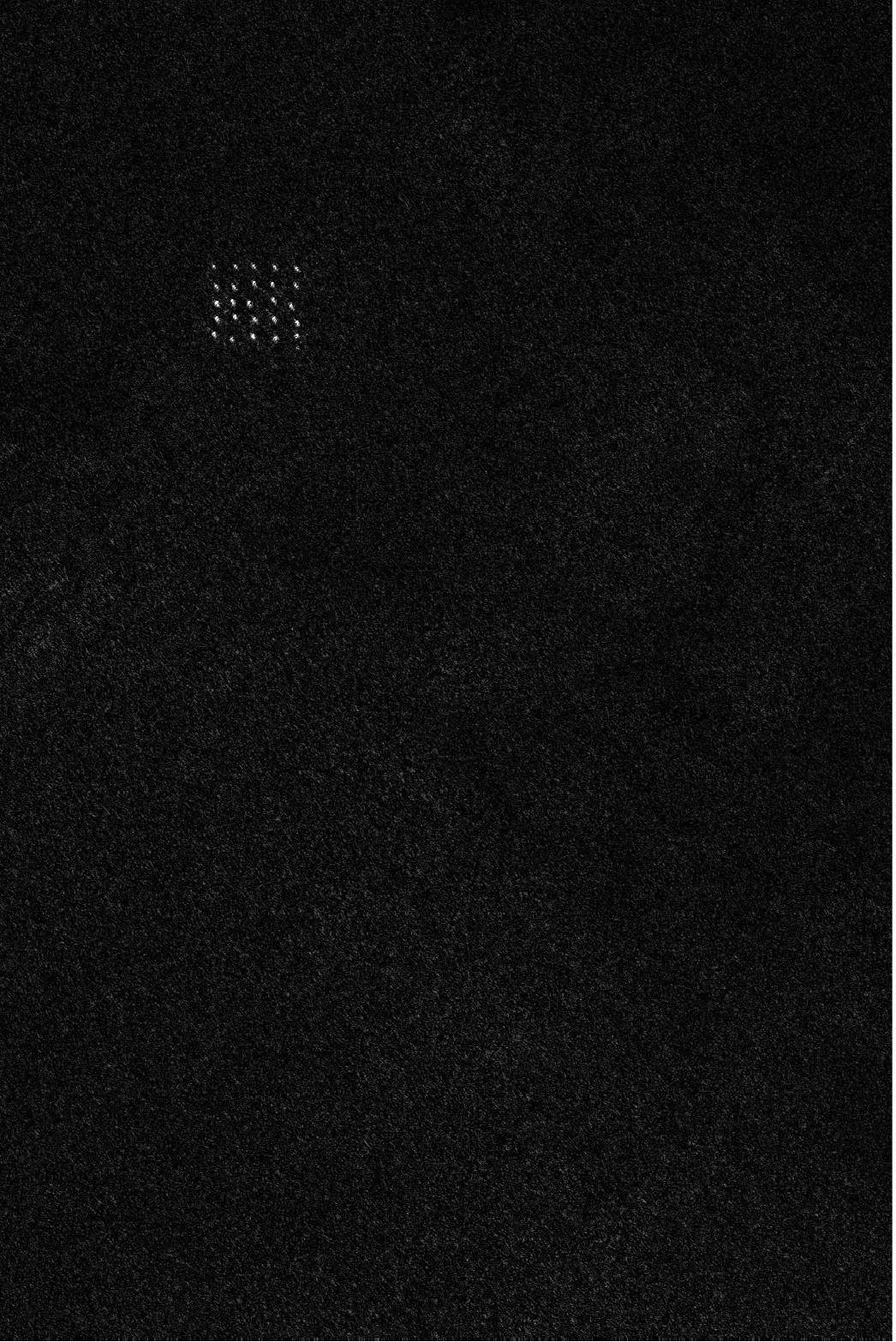}}
\subfigure[Mission two]
{\label{f:res2}\includegraphics[width=0.2\textwidth]{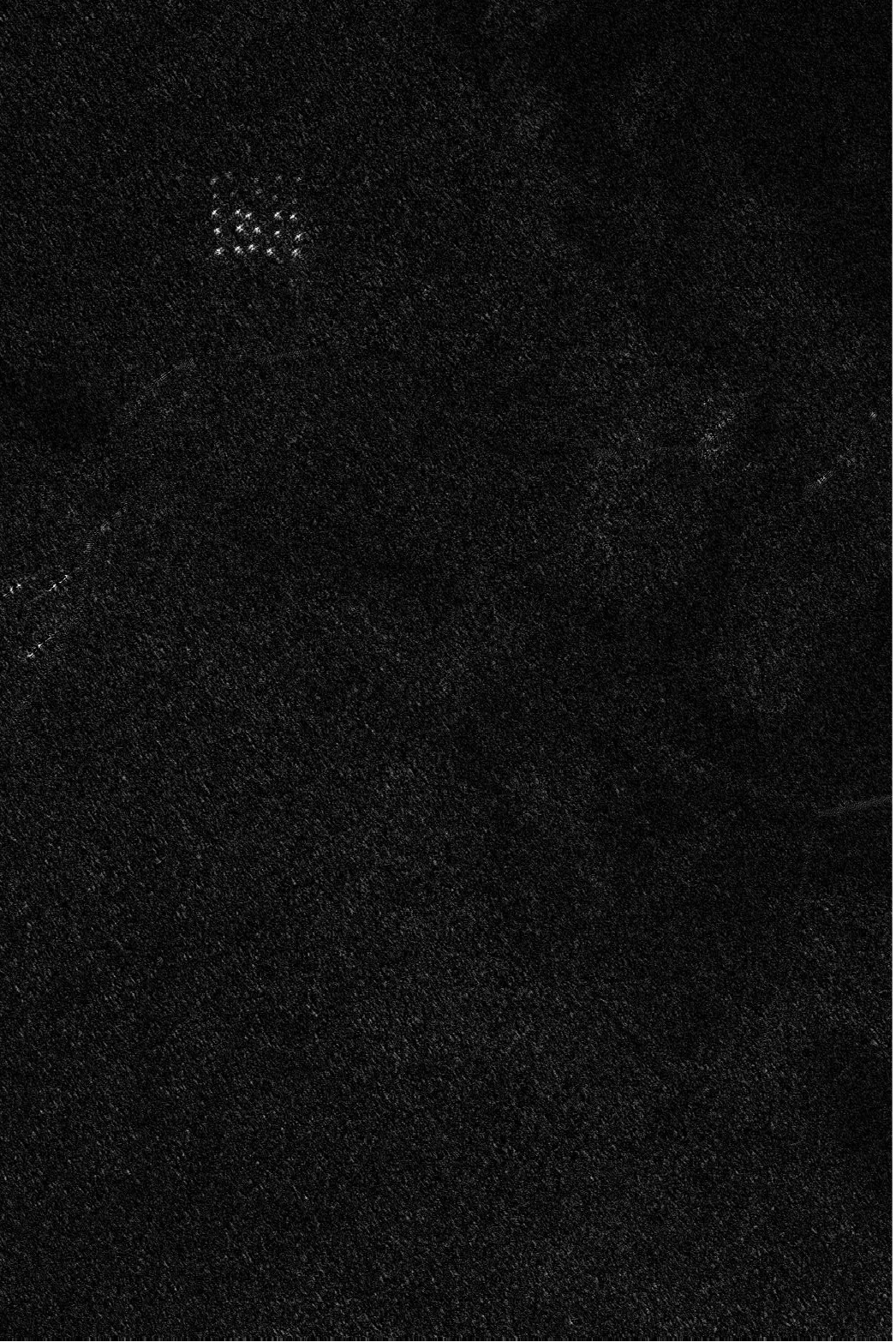}}
\subfigure[Mission three]
{\label{f:res3}\includegraphics[width=0.2\textwidth]{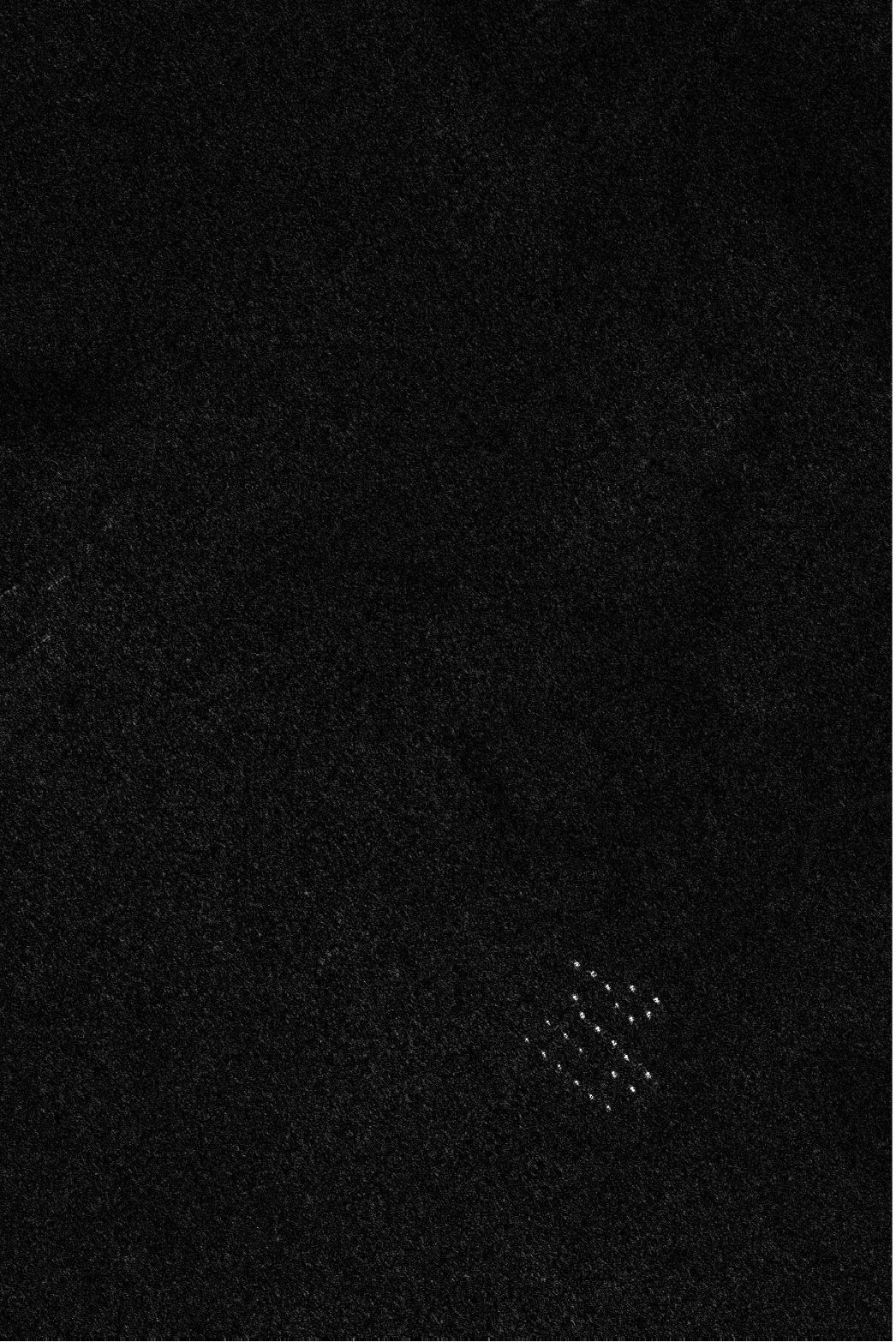}}
\subfigure[Mission four]
{\label{f:res4}\includegraphics[width=0.2\textwidth]{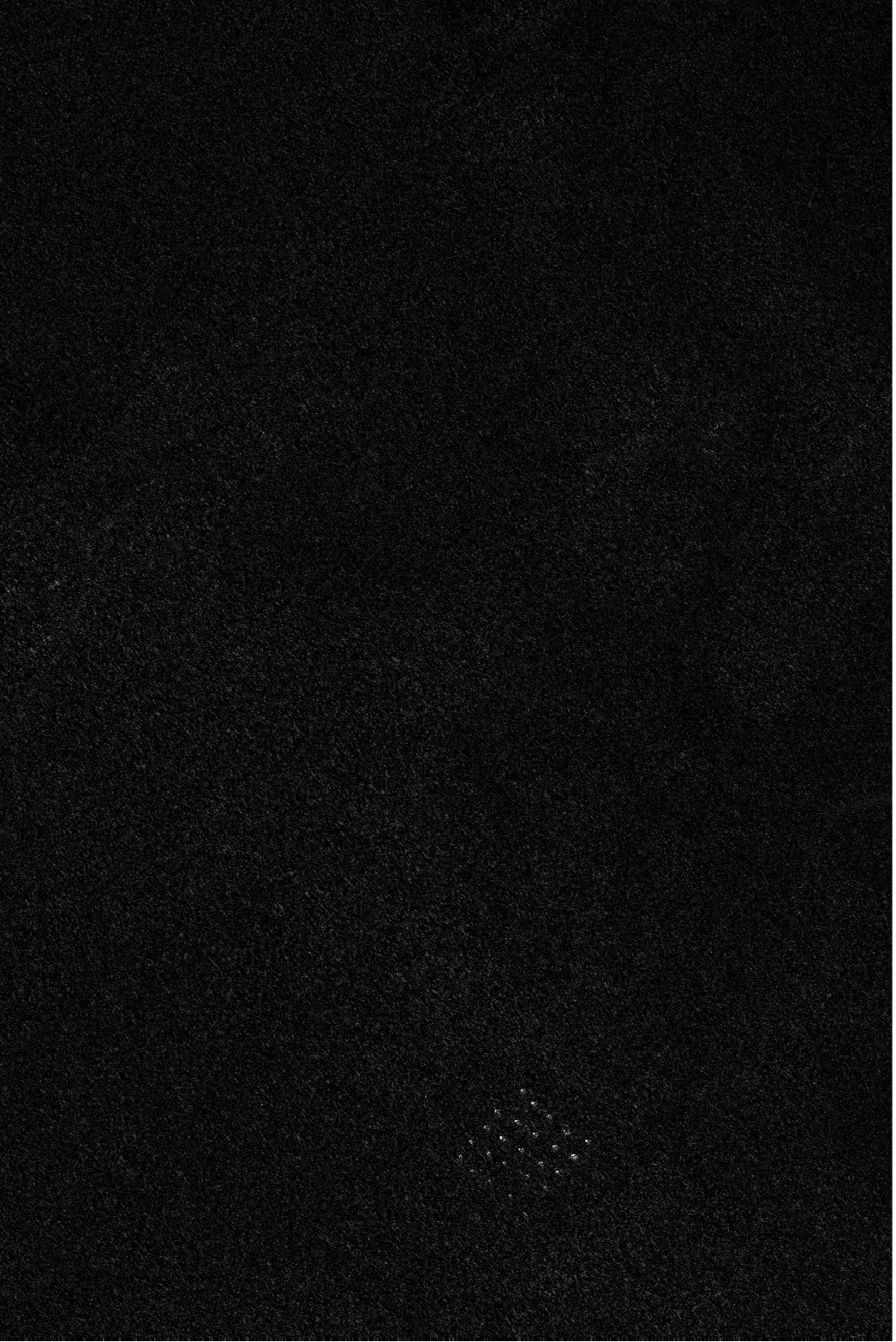}}
\caption{Difference images for pass five.}\label{f:rpass5}
\end{figure}

\begin{figure}
\centering
\subfigure[Mission one]
{\label{f:res5}\includegraphics[width=0.2\textwidth]{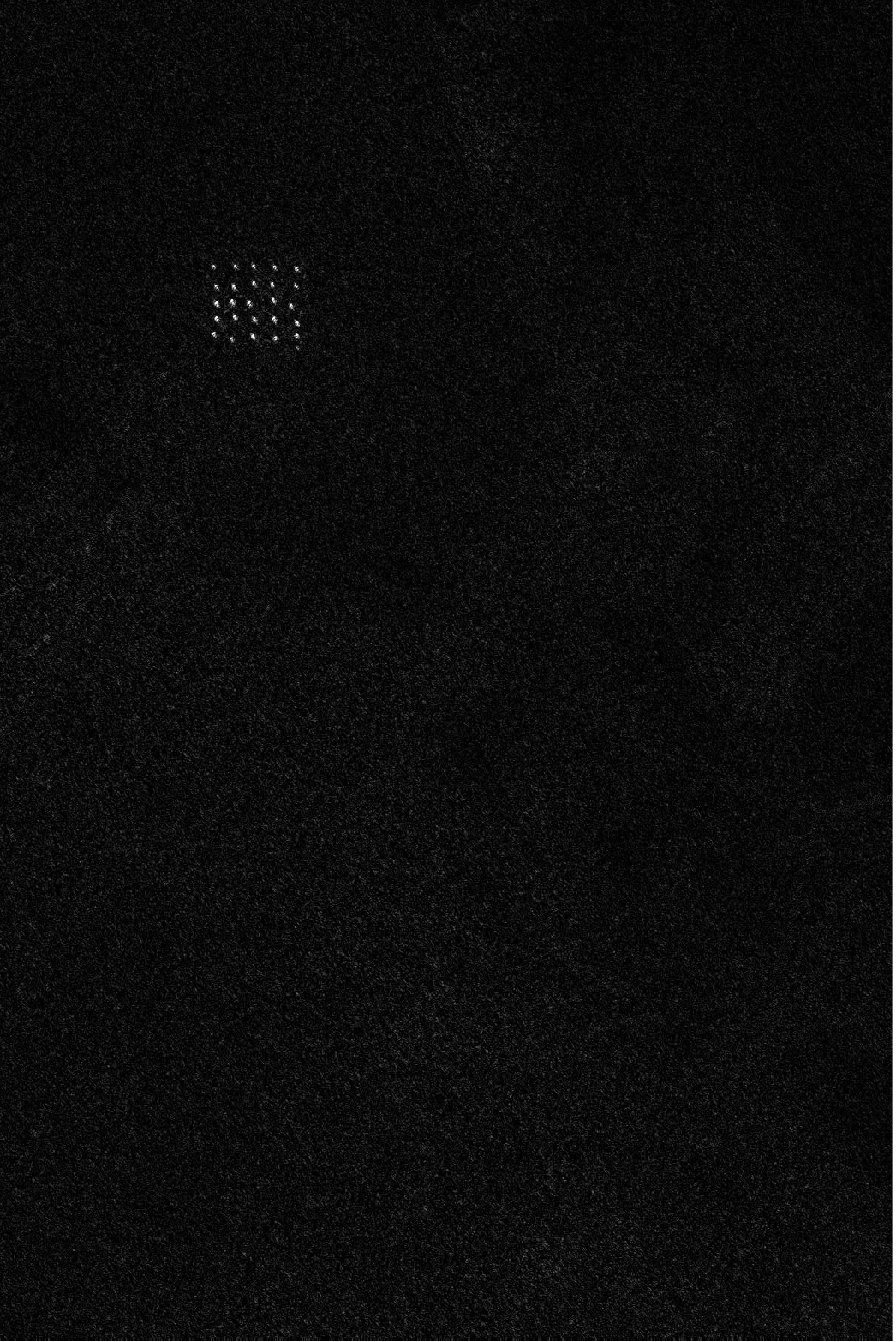}}
\subfigure[Mission two]
{\label{f:res6}\includegraphics[width=0.2\textwidth]{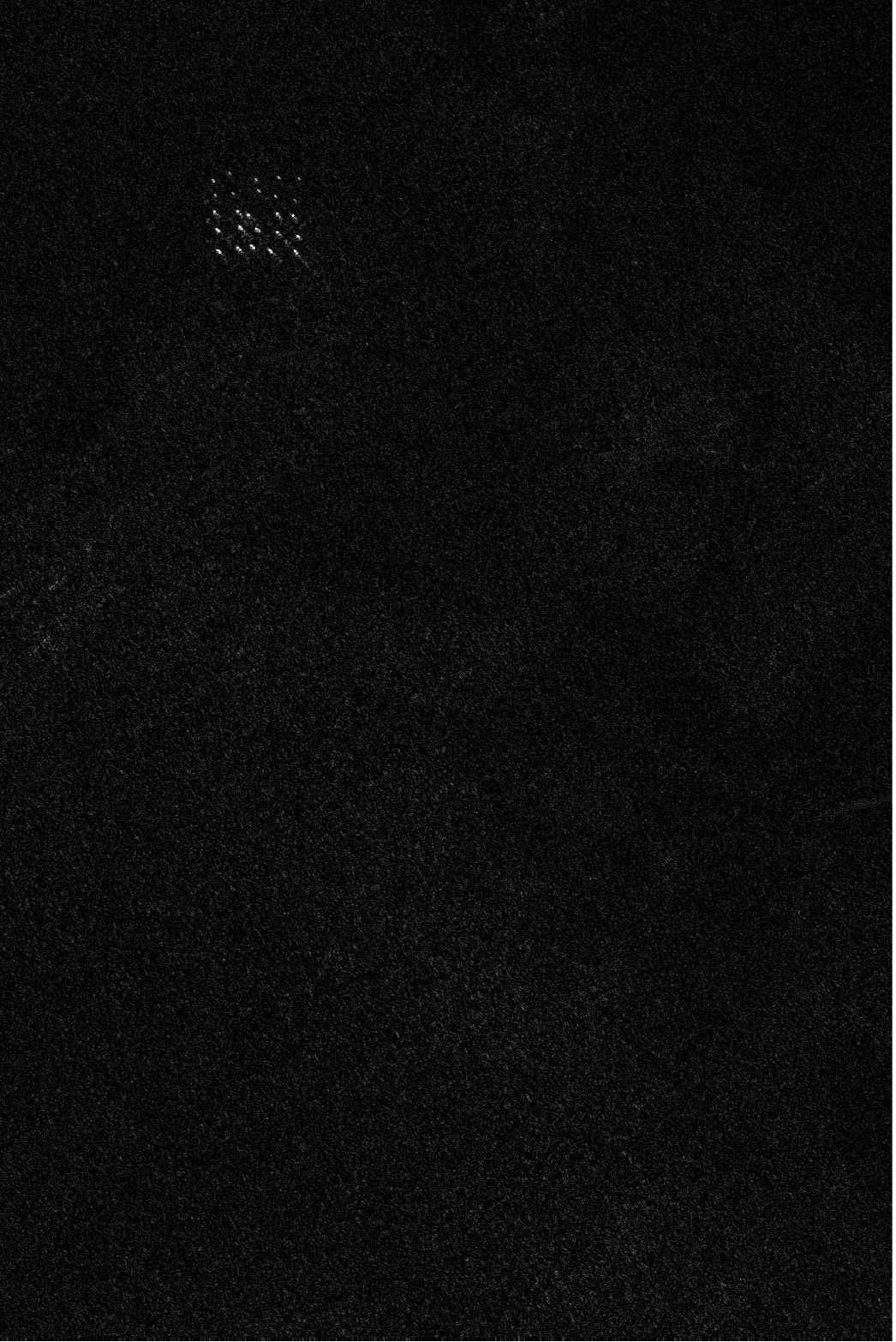}}
\subfigure[Mission three]
{\label{f:res7}\includegraphics[width=0.2\textwidth]{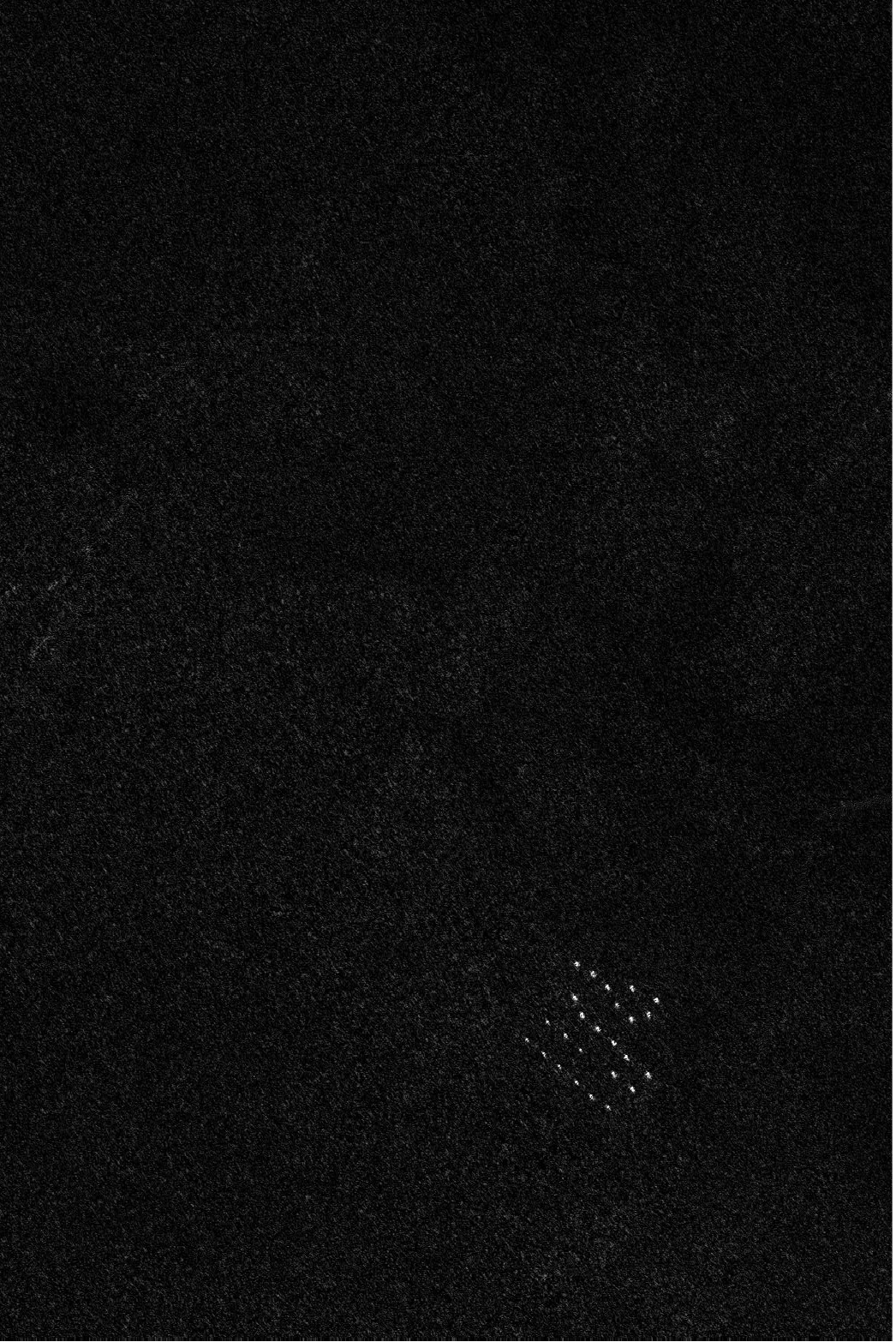}}
\subfigure[Mission four]
{\label{f:res8}\includegraphics[width=0.2\textwidth]{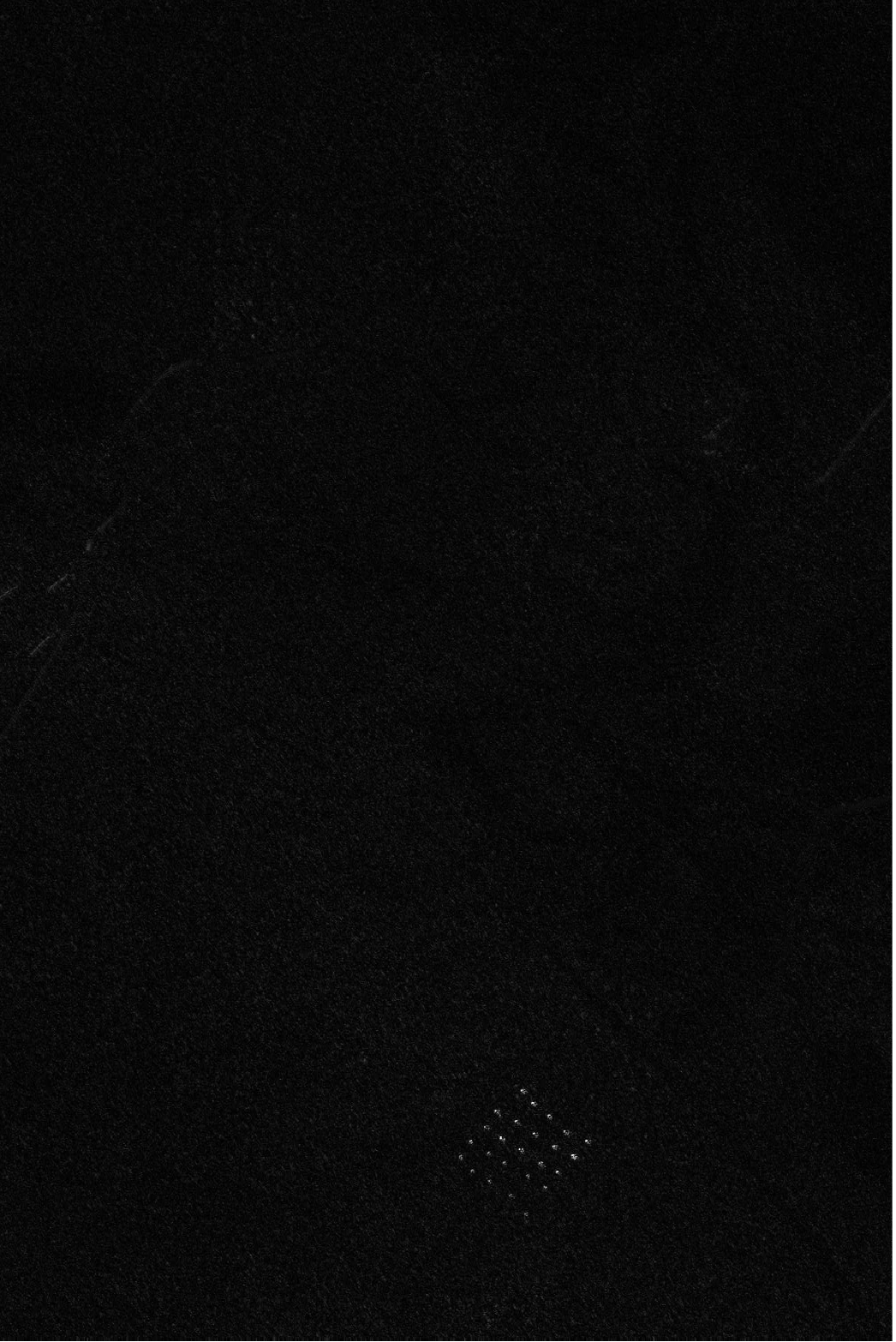}}
\caption{Difference images for pass six.}\label{f:rpass6}
\end{figure}

Figures~\ref{f:rpass5}
and~\ref{f:rpass6}
present
the images
obtained
by the
subtraction
between the
image of interest
and
the ground
estimation image
obtained by the AR$(1)$ model.
We can see that the
deployed vehicles are
more highlighted
in
Figures~\ref{f:rpass5}
and~\ref{f:rpass6}.
Only in Figure~\ref{f:res2}
we can verify
the strong
scatters
in the greater prominence.
In all other images,
the strong scatters
were almost completely
removed.

\subsection{Application in change detection}

\begin{figure}
\centering
\includegraphics[width=0.7\textwidth]{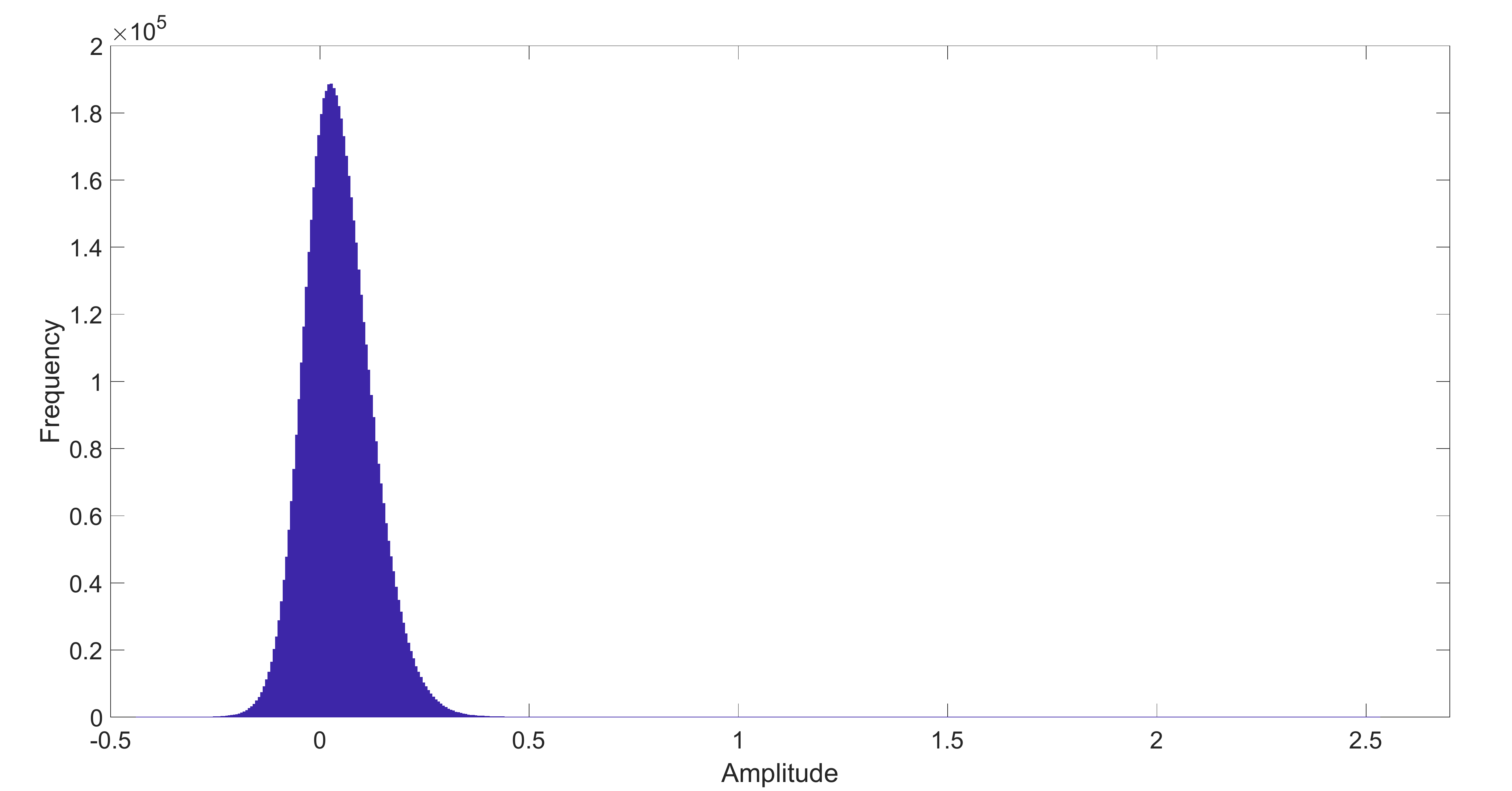}
\caption{Histogram for Figure~\ref{f:res1}.}\label{f:hist}
\end{figure}

We consider CDA as an application to the obtained ground estimation.
The change detection is realized
according to the processing scheme presented
in Figure~\ref{f:proc}. The AR model gives the ground estimation
presented in Figure~\ref{f:prev}.
With the estimated image of the
ground scene, we performed subtractions between the interest
images and the obtained images as given in
Figures~\ref{f:rpass5} and~\ref{f:rpass6}.
The next step is reserved for thresholding and
morphological operations to obtain the change detection
results.
Figure~\ref{f:hist} presents the histogram for
Figure~\ref{f:res1}.
We can see that the magnitude distribution of the images
obtained by the subtraction between the interest image and the
ground estimation image is approximately a normal distribution.

Thus,
the
thresholding
can be choosing
as
\begin{align*}
C = \frac{\lambda - \widehat{\mu}}{\widehat{\sigma}},
\end{align*}
where~$C$ is the detection constant,
$\widehat{\mu}$~is the mean,
and
$\widehat{\sigma}$
the
standard deviation of the
considered
pixels
in the image stack.
For evaluation, the values
$4.5; 5; 5.5; 6; 6.5$
are selected for~$C$.

Table~\ref{t:results}
presents the
change detection results
for~$C=4.5$.
We detect~$188$ deployed vehicles with~$33$
false alarms, i.e.,
the detection probability is~$94\%$
while the false alarm rate is only~$0.69/\text{km}^2$.
In these same images, in~\cite{Lundberg2006},
it was
detected~$188$ vehicles with~$52$ false alarms.
Hence,
considering the
AR model
to estimate the ground scene,
we obtained the same
number of detected
vehicles with less~$19$ false alarms than~\cite{Lundberg2006}.

Especially, the change detection results for mission
four and pass five is two false
alarms in comparison
to~$29$ in~\cite{Lundberg2006} (one less detection).
Among the $33$ false alarms presented in
Table~\ref{t:results},
$24$ of them
are related to
the images in
Figures~\ref{f:res2}
and~\ref{f:res8}.
Other
approaches to ground estimation and other
values of~$p$
can be further investigated to provide more accurate predicted SAR images.

\begin{figure}
\centering
\includegraphics[width=0.7\textwidth]{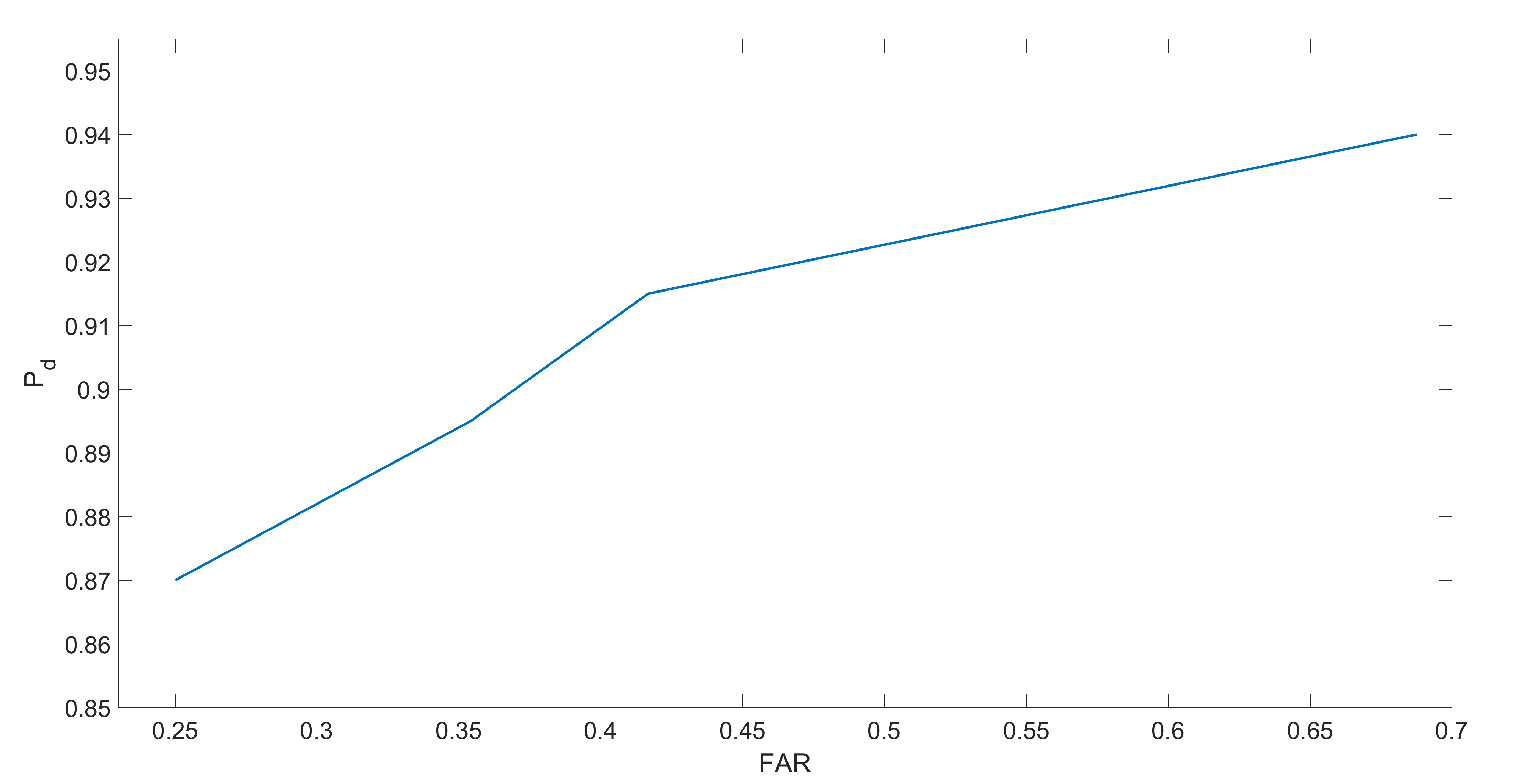}
\caption{ROC curve obtained with the proposed method,
for $C_i$, $i = 4.5,\,5,\,5.5,\,6,\,6.5$.}\label{f:roc}
\end{figure}

Figure~\ref{f:roc} presents the
ROC curve~\cite{ROC}
of the change detection results,
showing the probability of detection
versus the false alarm rate
for different values of threshold.
We can observe in the ROC
curve values of
$\text{P}_d =0.92$ with $\text{FAR} = 0.42$
or
$\text{P}_d = 0.90$ with $\text{FAR} = 0.35$,
as examples.
\begin{table}
\centering
\footnotesize
\caption{Change detection results obtained with~$C = 4.5$}
\label{t:results}
\begin{tabular}{cccccccc}
\hline
\multicolumn{2}{c}{Case of Interest} & Number of &
Detected & $\text{P}_d$ & Area &	Number of & FAR \\
Mission & Pass & known targets & Targets &  & $[\text{Km}^2]$ & false alarms \\
\hline
$1$	& $5$ &	$25$ &	$25$ &	$1.00$ & $6$ & $0$ & $0.00$ \\
$2$	& $5$ &	$25$ &	$16$ &	$0.64$ & $6$ & $9$ & $1.50$ \\
$3$	& $5$ &	$25$ &	$25$ &	$1.00$ & $6$ & $1$ & $0.17$ \\
$4$	& $5$ &	$25$ &	$22$ &	$0.88$ & $6$ & $2$ & $0.33$ \\
$1$	& $6$ &	$25$ &	$25$ &	$1.00$ & $6$ & $1$ & $0.17$ \\
$2$	& $6$ &	$25$ &	$25$ &	$1.00$ & $6$ & $2$ & $0.33$ \\
$3$	& $6$ &	$25$ &	$25$ &	$1.00$ & $6$ & $3$ & $0.50$ \\
$4$	& $6$ &	$25$ &	$25$ &	$1.00$ & $6$ & $15$ & $2.50$ \\
\hline
\multicolumn{2}{c}{Total}	& $200$ & $188$ & $0.94$ & $48$ & $33$ & $0.69$\\
\hline
\end{tabular}
\end{table}

\section{Conclusion}
\label{s:con}

Usually, AR models
are used in
the study
of
data that are defined in
the time domain.
In this paper we proposed the use of an
AR$(1)$ model for
a stack of eight SAR images
to retrieve a
ground scene estimation.
By using this technique, it was
possible to obtain a reliable representation
of the ground scene,
which could be used
as
a reference image in a
change detection algorithm.
We applied
subtractions
between the
ground estimation image
obtained by
AR models
and the interest images.
With
simple
thresholding and morphological operations,
we obtained
competitive results
of~$\text{P}_d$
and FAR
when compared
to the results presented in~\cite{Lundberg2006}.

To the best of our knowledge,
this is
the first
treatment
for
ground estimation
with
image stack
by using AR models.
In future studies,
we intend to use
stacks with more images
generating possibly more accurate adjustments
and consequently
more reliable
forecasts or more accurate ground estimation.

\section*{Acknowledgements}

We gratefully
acknowledge partial financial support from
Conselho Nacional de Desenvolvimento
Cient\'ifico, Tecnol\'ogico (CNPq),
and Coordena\c{c}\~ao de Aperfei\c{c}oamento
de Pessoal de N\'ivel Superior (CAPES),
and Swedish-Brazilian Research and Innovation Centre (CISB),
Brazil, and the Saab AB company.
We also would like thank
the Swedish Defence Research
Agency (FOI) for
providing the CARABAS
SAR data.

{\small
\singlespacing
\bibliographystyle{siam}
\bibliography{bib}
}

\end{document}